\documentclass[pre,10pt,twocolumn,groupedaddress,showpacs,floatfix,superscriptaddress]{revtex4}
\usepackage[dvips]{graphicx}
\usepackage{dcolumn}
\usepackage{amsmath}
\usepackage{bm}
\begin{document}

\title{Forces between elongated particles in a nematic colloid}

\author{D.~Andrienko}
\affiliation{Max Planck Institute for Polymer Research, Ackermannweg 10, 55128 Mainz, Germany}

\author{M.~Tasinkevych}
\affiliation{Departamento de F\'\i sica da Faculdade de Ci\^encias and Centro de F\'\i sica Te\'oretica e Computacional, Univercidade de Lisboa, Avenida Professor Gama Pinto 2, P-1649-003 Lisboa Codex, Portugal}

\author{P.~Patr{\'\i}cio}
\affiliation{Departamento de F\'\i sica da Faculdade de Ci\^encias and Centro de F\'\i sica Te\'oretica e Computacional, Univercidade de Lisboa, Avenida Professor Gama Pinto 2, P-1649-003 Lisboa Codex, Portugal}
\affiliation{Instituto Superior de Engenharia de Lisboa, 
Rua Conselheiro Em\'\i dio Navarro 1, P-1949-014 Lisboa, Portugal}

\author{M.~P.~Allen}
\affiliation{Center for Scientific Computing, University of Warwick, Coventry, United Kingdom}

\author{M.~M.~Telo da Gama}
\affiliation{Departamento de F\'\i sica da Faculdade de Ci\^encias and Centro de F\'\i sica Te\'oretica e Computacional, Univercidade de Lisboa, Avenida Professor Gama Pinto 2, P-1649-003 Lisboa Codex, Portugal}

\date{\today}
\begin{abstract}
Using molecular dynamics (MD) simulations we study the interactions between elongated colloidal particles (length to breath ratio $\gg 1$ ) in a nematic host. The simulation results are compared to the results of a Landau-de Gennes elastic free energy. We find that depletion forces dominate for the sizes of the colloidal particles studied. The tangential component of the force, however, allows us to resolve the elastic contribution to the total interaction. We find that this contribution differs from the quadrupolar interaction predicted at large separations. The difference is due to the presence of nonlinear effects, namely the change in the positions and structure of the defects and their annihilation at small separations.
\end{abstract}
\pacs{61.30.Cz, 61.30.Jf, 61.20.Ja, 07.05.Tp}
\maketitle

\section{Introduction}
Liquid crystal colloids belong to a special class of colloidal systems. Long-range orientational order of the liquid crystal molecules gives rise to additional {\em long-range} interactions between the colloidal particles \cite{poulin.p:1997.a,lubensky.tc:1998.a}. 
The presence of defects, due to topological restrictions, controls the symmetry of this interaction, that may be of dipolar or quadrupolar symmetry \cite{stark.h:2001.a,mondain-monval.o:1999.a,gu.y:2001.a}. 
Clustering, superstructures, and new phases are immediate consequences of additional anisotropic interactions \cite{meeker.sp:2000.a,anderson.vj:2001.a,anderson.vj:2001.b}.

The long-range forces between particles of any shape can, in principle, be calculated using direct integration over the director field \cite{lev.bi:1999.a}. Ready-to-use expressions are available for spherical particles \cite{ramaswamy.s:1996.a}, two-dimensional disks \cite{tasinkevych.m:2002.a}, etc. For smaller particle-particle separations nonlinear effects from the elastic free energy come into play. Minimization of the Landau-de Gennes free energy with respect to the tensor order parameter can be used to take into account the relative position of the defects and the variation of the nematic order parameter around the colloids \cite{tasinkevych.m:2002.a}. For even smaller separations, depletion forces, density variation, and presmectic ordering of a nematic liquid crystal near colloidal particles cannot be ignored. Then the density functional approach \cite{teixeira.pic:1997.a,allen.mp:2000.c,andrienko.d:2002.a} or, alternatively, computer simulation techniques \cite{billeter.j.l:2000.a,andrienko.d:2001.b,kim.eb:2002.a} can be used.

In this paper we study the interaction between elongated colloidal particles suspended in a nematic liquid crystal. The liquid crystal molecules are taken to be \emph{homeotropically anchored}, that is, their preferred orientation is normal to the colloid surface. We compare the results of two methods: molecular dynamics simulation and minimization of the phenomenological Landau-de Gennes free energy. 

It has already been shown theoretically \cite{burylov.sv:1994.a} and using MD simulation \cite{andrienko.d:2002.a} that isolated small elongated particles, with homeotropic boundary conditions, minimize the free energy by orienting perpendicularly to the director. Thus, we consider the particle symmetry axes parallel to each other and perpendicular to the far field director. In this geometry, the defect structure around a single particle has been studied in detail \cite{andrienko.da:2002.b}. The configuration with two $1/2$ strength defects is preferable energetically, giving rise to a quadrupolar-type of interaction between two such particles \cite{ramaswamy.s:1996.a}. Analytical expressions for the {\em long-range} forces derived from the Frank elastic free energy exist, as well as numerical studies based on the tensor order parameter formalism \cite{fukuda.j:2001.a,tasinkevych.m:2002.a}.

However, more detailed studies of the forces showed that the interaction between the particles is no longer quadrupolar at small separations~\cite{tasinkevych.m:2002.a}, due to a change in the relative position of the defects. It was shown that the long-range repulsive interactions can become attractive for small anchoring strengths while remaining repulsive for all orientations for strong anchoring. As the distance between the particles decreases, their preferred relative orientation with respect to the far field nematic director changes from oblique ($\pi/4$ for the pure quadrupolar interaction) to perpendicular.

In this paper we use MD simulations to confirm this conclusion at even smaller separations, where the depletion forces play an important role. To resolve the elastic contribution to the total force we measure separately the {\em normal} and the {\em tangential} component of the force. 
The normal component is much larger than the tangential component and it is practically unaffected by the relative position of the defects. The tangential component, however, has a dependence on the particle separation that is qualitatively the same as that predicted by the minimization of the Landau-de Gennes free energy.

The paper is organized as follows. In Sec.~\ref{sec:model} we describe the geometry, the molecular model used for MD simulation, and the technique of minimization with adaptive meshing. The director orientation, order parameter, and density maps, as well as forces between the particles are presented in Sec.~\ref{sec:results}. In Sec.~\ref{sec:conclusions} we make some concluding remarks.

\section{Molecular model and simulation methods}
\label{sec:model}
\begin{figure}
\begin{center}
\includegraphics[width=5cm, angle=0]{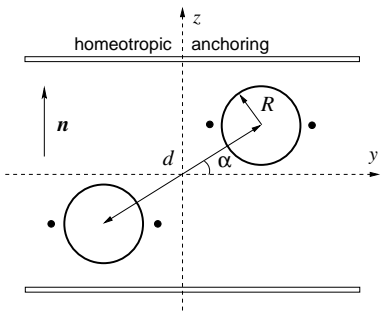}
\end{center}
\caption[Studied geometry]{
\label{fig:1} 
 Geometry ($yz$ cross-section is shown): two cylinders of radius $R$ are immersed in a liquid crystal host. The host is modeled as a solution of Gay-Berne particles and the orientation of the director is fixed at the top and bottom walls of the simulation box ($z = \pm L_z/2$). Periodic boundary conditions are applied in the $x,y$ directions.}
\end{figure}

The geometry considered in this work is shown in Figure \ref{fig:1}. Two cylindrically shaped colloidal particles of radius $R$ are immersed in a liquid crystal. The particles are separated by a distance $d$ measured from their symmetry axes.  The director orientation is fixed at the top and the bottom walls, parallel to the $z$ axis and perpendicular to the symmetry axes of the cylinders. Boundary conditions ensure that the director $\bm n$ far from the colloidal particles is parallel to the $z$ axis. The rod length is considered to be infinite: in the simulations this simply means that the cylinder spans the $x$-dimension of the periodic box.

\subsection{MD simulation}
Molecular dynamics simulations were carried out using a soft repulsive potential, describing (approximately) ellipsoidal molecules of elongation $\kappa = 3$ \cite{andrienko.d:2001.b}. The systems consisted of $N = 8,000$ and $64,000$ particles (see table~\ref{tab:1} for details). A reduced temperature $k_{\rm B}T/\epsilon_0=1$ was used throughout. The system size was chosen so that the number density is $\rho\sigma_0^3 = 0.32$, within the nematic phase for this system. Here $\sigma_0$ is a size parameter, $\epsilon_0$ is an energy parameter (both taken to be unity).

\begin{table*}
\caption[Studied systems]{
\label{tab:1}
Systems studied}
\begin{ruledtabular}
\begin{tabular}{cccc}
Colloid radius & Particles & Box size & Production \\
$R/\sigma_0$    & $N$  & $x/\sigma_0, y/\sigma_0, z/\sigma_0$ & (steps, $10^5$)\\
\hline
 3 & 8,000 & 10, 50, 50 & 10 \\
 5 & 64,000 & 20, 100, 100 & 5 \\
\end{tabular}
\end{ruledtabular}
\end{table*}

 The system is confined in the $z$ direction, to provide uniform orientation of the director far from the particles along the $z$ axis. The interaction of molecule $i$ with the colloidal particle (rod) and the wall was given by a shifted Lennard-Jones repulsion potential having exactly the same form as in Refs.~\cite{andrienko.d:2001.b,andrienko.da:2002.b}. This provides homeotropic orientation with a strong anchoring of the molecules at the wall.

\begin{figure}
\begin{center}
\includegraphics[width=8cm, angle = 0]{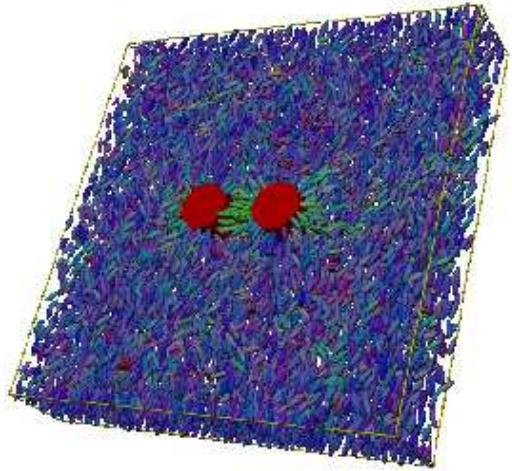}
\caption[]
{MD simulation results: snapshot of the system. Number of particles $N = 8000$, colloid radius $R/\sigma_0 = 3$, colloid separation $d/\sigma_0 = 10$, $\alpha = 0$. Color coding emphasizes particle orientations.}
\label{fig:md_snapshot}
\end{center}
\end{figure}

The radius and the length of the rod were steadily increased from zero to the desired value during $10^3$ steps. Then the system was equilibrated for $10^5$ steps. During equilibration we scaled the velocities of the molecules to achieve $k_{\rm B}T/\epsilon_0 = 1$. An equilibrated snapshot of the system is shown in Fig.~\ref{fig:md_snapshot}.

The production run for every angle $\alpha$ was $10^6$ steps. The force ${\bm F}$ on the rod was calculated using the repulsive force ${\bm f}_i$ from the rod on the particle $i$
\begin{eqnarray}
{\bm F} &=& - \sum_{i=1}^N  {\bm f}_{i}.
\label{torque}
\end{eqnarray}

The local tensor order parameter $\bm{Q}(\bm{r})$ was calculated as
\begin{equation}
\label{order_tensor}
Q_{\alpha \beta}(z_{i},y_{j}) = \frac{1}{n_{\{i,j\}}} \sum_{k=1}^{n_{\{i,j\}}}
\left\{ \frac{3}{2}\left<u_{k \alpha}u_{k \beta}\right> - 
\frac{1}{2} \delta_{\alpha \beta} \right\},
\end{equation} 
where there are $n_{\{i,j\}}$ molecules present in each bin ${\{i,j\}}$, $\delta_{\alpha \beta}$ is the Kronecker delta, $\left< \cdots \right>$ denotes an ensemble average, $\alpha, \beta = x, y, z$. Diagonalizing the $Q_{\alpha\beta}$ tensor, for each bin, gives three eigenvalues, $Q_1$, $Q_2$, and $Q_3$, plus the three corresponding eigenvectors. The eigenvalue with the largest absolute value defines the order parameter $S$ for each bin. The biaxiality $B$ is calculated as the absolute value of the difference between the remaining two eigenvalues of the tensor order parameter.

\subsection{Landau-de Gennes free energy minimization}
In the framework of the continuum theory, the system can be described by the Landau-de Gennes free energy~\cite{degennes.pg:1995.a}
\begin{equation}
{\cal F}\{\bm Q\} = \int_\Omega \left[
\frac{L}{2} \left| \nabla {\bm Q} \right| ^{2}
+f \left({\bm Q} \right)
\right],
\label{free_energy}
\end{equation}
where $f\left(\bm Q\right)$ is a function of the invariants of $\bm Q$, the symmetric tensor order parameter, and the integration extends over the sample volume. Here we adopted the one-constant approximation for the elastic free energy (decoupling of the spatial and spin rotations).

To simplify the calculations, we used a two-dimensional representation of the tensor order parameter. This representation is justified for a uniaxial nematic with the director in the $yz$ plane. Indeed, as we will see, the MD simulation results show that the director is constrained in the $yz$ plane. The nematic is uniaxial in the bulk and is slightly biaxial in the defect core.

In this approach it is possible to rewrite $\bm Q$ in terms of components of the director ${\bm n}$ and the scalar order parameter $Q$
 \begin{equation}
Q_{ij} = Q \left( n_i n_j - \frac{1}{2}\delta_{ij}\right). 
\label{eq_tensor_definition}
\end{equation}

Then the Landau-de Gennes free energy density reads
\begin{equation}
f({\bm Q}) =  -\frac{a}{2} {\rm Tr}{\bm Q}^2 
+\frac{c}{4}\left[{\rm Tr}{\bm Q}^2\right]^2,
\label{free_energy_density}
\end{equation}
where $a$ is assumed to depend linearly on the temperature, whereas the positive constant $c$ is considered temperature independent. Note, that the invariant that corresponds to the cubic term of the tensor order parameter vanishes in the two-dimensional nematic.

For this free energy the liquid crystal in the nematic state has the order parameter $Q_{eq}=\sqrt{2a/c}$. The elastic constant $L$ is related to the Frank elastic constant by $K=4La/c$. In our calculations we used $a=1, c=2$ which gives $Q_{eq}=1$.   

The free energy ${\cal F}\{\bm Q\}$ was minimized numerically, using finite elements with adaptive meshes. The geometry used in the numerical calculation is shown schematically in figure \ref{fig:1}. During minimization we used a square integration region $\Omega$ of size $40R\times 40R$, where $R$ is the radius of the colloidal particle. The area $\Omega$ was triangulated using a $\rm BL2D$ subroutine~\cite{george.pl:1998.a}.
The tensor order parameter $\bm Q$ was set at all vertices of the mesh and was linearly interpolated within each triangle. Using standard numerical procedures the free energy was minimized under the constraints imposed by the boundary conditions, i.e., strong homeotropic anchoring at the particle perimeters and uniform alignment at the outer boundary.

Finally, a new adapted mesh was generated iteratively from the result of the 
previous minimization. The new local triangle size was related to the free energy variations of the previous solution, assuring a constant numerical weight for each minimization variable. The final meshes with a minimal length of $\sim 10^{-3}R$ had about $2\times 10^4$ minimization variables.

The tangential component of the interparticle force, $F_{\tau}$, was calculated numerically from the free energy, $F_{\tau}=-r^{-1}\partial {\cal F}/\partial \alpha$.
\section{Results}
\label{sec:results}

\subsection{Defect structures}
\label{sec:defects}

\begin{figure}
\begin{center}
\includegraphics[width=8cm]{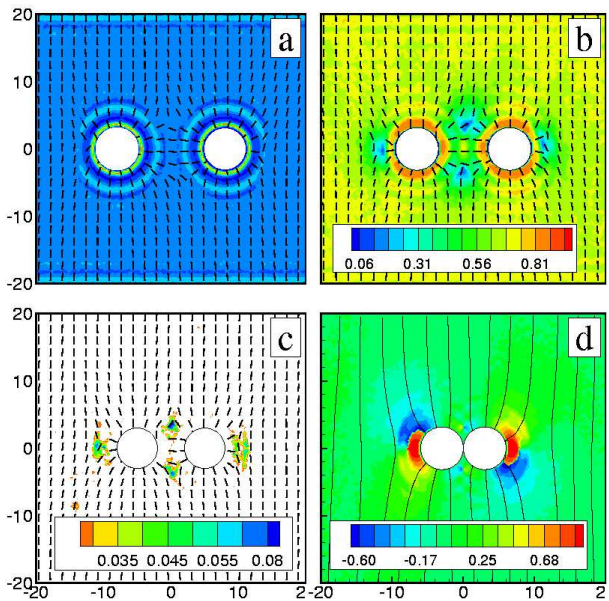}
\caption[]
{MD simulation results. Maps of (a) density at the separation $d / \sigma_0 = 16$, (b) order parameter at the separation $d / \sigma_0 = 13$ , and (c) biaxiality at the separation $d / \sigma_0 = 13$. (d) shows the map of the $n_y$ component of the director at the separation $d / \sigma_0 = 6$. The $yz$ cross section of the system is shown. Rod radius $R = 3\sigma_0$. The director far from the particle is constrained along the $z$ axis.}
\label{fig:md_density}
\end{center}
\end{figure}

A typical director orientation together with the density, order parameter, and biaxiality maps are shown in Fig.~\ref{fig:md_density}. For large separations, a pair of defect lines forms perpendicular to the director, next to each particle, and stays at the same positions as in the case with a single colloidal particle. The director distortion vanishes very quickly in the liquid crystal bulk, and the core region extends over a few molecular lengths~\cite{andrienko.da:2002.b}. 

On reducing the separation, at a certain value the positions of the defects start to change, as well as their inner structure (see Fig.~\ref{fig:md_density}b,c). Finally, when the particles are about to merge, two out of four defects vanish to ensure that the total topological charge in the system is zero (Fig.~\ref{fig:md_density}d).

This scenario agrees qualitatively with the results of the minimization of the Landau-de Gennes free energy~\cite{tasinkevych.m:2002.a}. Using MD simulation, we are also able to observe the annihilation of the defects, which is not accessible on length scales where the phenomenological approach is applicable. From the density map shown in Figure~\ref{fig:md_density}a one can already foresee the strong influence of depletion effects at small separations. The order parameter map (Fig.~\ref{fig:md_density}b) illustrates the dislocation of the defects and the changes in the defect cores.

\subsection{Force between the particles}
\begin{figure}
\begin{center}
\includegraphics[width=8cm, angle = 0]{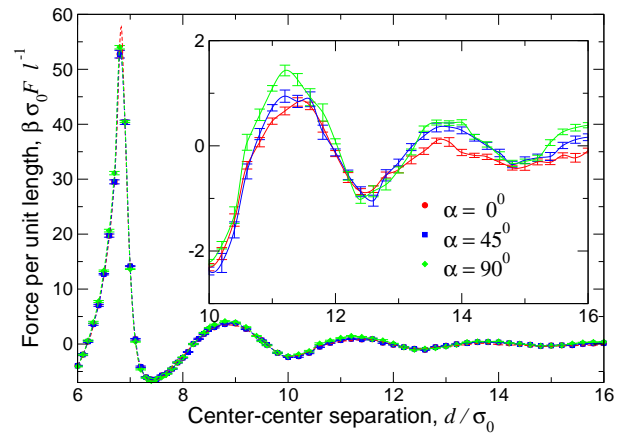}
\caption[]
{MD simulation results: components of the force parallel to the particle-particle separation vector, $F_\parallel$, as a function of the particle-particle separation $d$. Colloid radius $R/\sigma_0 = 3$.}
\label{fig:force_p_R3}
\end{center}
\end{figure}

Figure~\ref{fig:force_p_R3} shows typical force curves as a function of the particle-particle separation $d$ for colloidal particles of size $R/\sigma_0 = 3$. From the parallel component of the force (projection of the force on the particle-particle separation vector) one can see that the depletion forces indeed dominate; the force has oscillations due to the density modulation of the liquid crystal close to the particle surfaces (presmectic ordering). More detailed analysis of the force curves, shown in the inset of Fig.~\ref{fig:force_p_R3}, reveals that there is a small difference in the decaying tails of the force curves, which can be attributed to the elastic contribution of the order parameter field/defects around the colloidal particles to the total interparticle interaction. 

\begin{figure}
\begin{center}
\includegraphics[width=8cm, angle = 0]{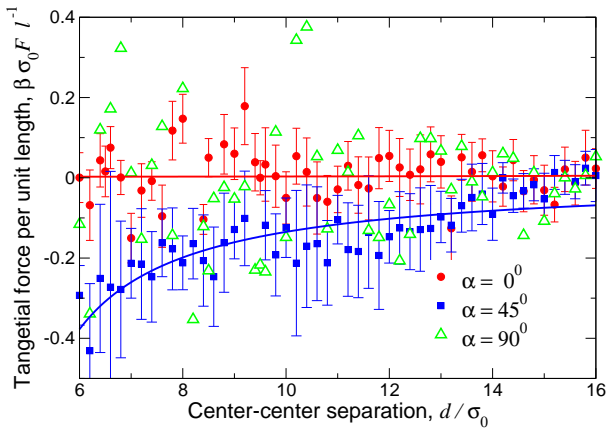}
\caption[]
{MD simulation results: component of the force perpendicular to the particle-particle separation vector, $ F_\tau$, as a function of the particle-particle separation $d$. Colloid radius $R/\sigma_0 = 3$. Error bars for $\alpha=\pi/2$ are too big to show on the plot. Smooth curves are to guide the eye.}
\label{fig:force_t_R3}
\end{center}
\end{figure}

To emphasize the contribution of the elastic force, we plot the {\em tangential} component of the force (perpendicular to the particle-particle separation vector) in Fig.~\ref{fig:force_t_R3}. We first look at the situation with $\alpha = \pi/4$. It is clear that there is a non-zero tangential component, which decays with the distance, i.e.\ there is a non-zero force which tends to align the particle-particle separation vector perpendicular to the director (or at some angle which is less than $\pi/4$). This is already different from the prediction of the quadrupolar interaction, where the minimum of the free energy is at $\alpha = \pi/4$, and agrees with the results of the Landau-de Gennes free energy minimization for small particle separations, see Ref.~\cite{tasinkevych.m:2002.a}. 

The situation with $\alpha = 0, \pi/2$ is somewhat different: there is a scatter in the value of the tangential component of the force, accompanied with a large error in the measurements. Analysis of the configurations suggests that this is due to the degeneracy in the possible defect positions. Indeed, if the particles move close to each other, the defects change their positions. The vector between the defects belonging to the same particle tilts with respect to the director. If $\alpha = 0$ or $\pi/2$, the tilt angle can be either positive or negative: both configurations are equivalent and have the same energy. However, there is a barrier between these two configurations. For small particles, this barrier is of the order of $k_B T$, and the defects can switch between two equivalent configurations during the simulation run. This `drift' of the defects does not affect the parallel component of the force, but leads to the `averaging' of the tangential component to zero. In addition, this drift is rather slow on the timescale of a molecular simulation, and provides a large scatter in the value of the tangential force.

\begin{figure}
\begin{center}
\includegraphics[width=8cm, angle = 0]{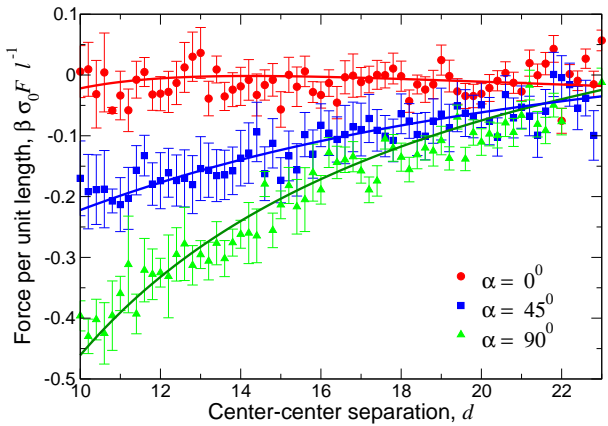}
\caption[]
{MD simulation results: component of the force perpendicular to the particle-particle separation vector, $F_\tau$, as a function of the particle-particle separation $d$. Colloid radius $R/\sigma_0 = 5$. Smooth curves are to guide the eye.}
\label{fig:force_t_R5}
\end{center}
\end{figure}

To overcome this problem, we also studied particles with a bigger radius, $R = 5 \sigma_0$. The elastic contribution to the force is larger in this case. As a consequence, the energy barrier between the two equivalent configurations is also larger. It is harder for the defects to overcome this barrier, which becomes larger than $k_BT$ for this size of the colloids. This is clearly seen from the tangential component of the force plotted in Fig.~\ref{fig:force_t_R5}; the scatter of the data is much smaller and there is a clear decrease with the decrease of the particle-particle angle. One can also see that there is a non-zero tangential force for $\alpha=\pi/2, \pi/4$, i.e.\ the particles tend to align in a way that their center-center separation vector is perpendicular to the director.

\begin{figure}
\begin{center}
\bigskip
\includegraphics[width=8cm]{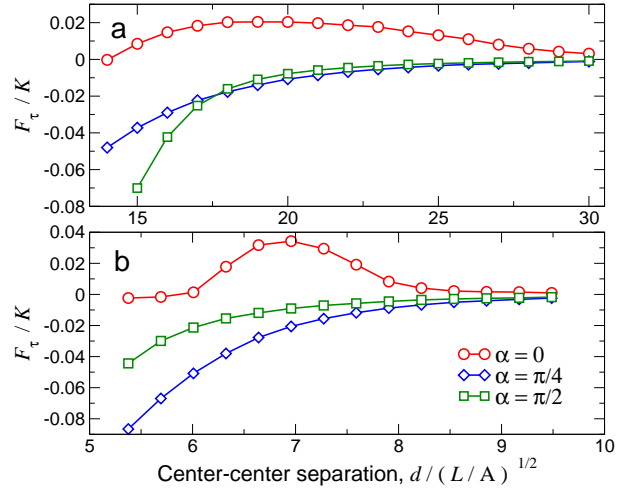}
\caption[]
{Landau-de Gennes theory: component of the force perpendicular to the vector joining the centers of the colloids as a function of the colloid separation, $d$, at three different orientations ($\alpha=0, \pi/4, \pi/2$). $d$ is in units of the nematic correlation length $\zeta = \sqrt{L/a}$. a) $\zeta / R = 0.2$; b) $\zeta / R = 0.633$}
\label{fig:f_tan_landau}
\end{center}
\end{figure}

Finally, we show the results of the Landau-de Gennes theory in Fig.~\ref{fig:f_tan_landau}. The tangential component of the force is plotted as a function of the interparticle distance $d$, for three different orientations of the particle-particle separation vector, $\alpha=0, \pi/4, \pi/2$.

Comparing the results of the Landau-de Gennes theory, Fig.~\ref{fig:f_tan_landau}, to the MD simulation results, Fig.~\ref{fig:force_t_R5}, one sees qualitative agreement for $\zeta / R = 0.2$ where $\zeta = \sqrt{L/a}$ is the nematic correlation length.

\section{conclusions}
\label{sec:conclusions}
We used two independent techniques, molecular dynamics simulation and minimization of the Landau-de Gennes free energy, to study the interaction of two elongated colloidal particles embedded in a nematic host. Our results show that the particle-particle interaction is no longer quadrupolar at short distances due to a change in the relative position of the defects. MD simulation results also show that for small particles the depletion force dominates but contributes mostly to the interparticle radial force. The tangential contribution to the force is of elastic origin. Its dependence on the center-center separation is in qualitative agreement with the results of the free energy minimization.

\begin{acknowledgments}
MD simulations used the GBMEGA program of the `Complex Fluids Consortium'. DA acknowledges the support of the Alexander von Humboldt foundation. The support of the Funda\c c\~ao para a Ci\^encia e Tecnologia (FCT) through a grant (Programa Plurianual) and grants No. SFRH/BPD/1599/2000 (MT) and No. SFRH/BPD/5664/2001 (PP) is aknowledged.
\end{acknowledgments}

\bibliography{journals,extra}

\end{document}